\documentclass[11pt]{article}
\usepackage{amssymb}
\usepackage{amsmath}
\usepackage{amsthm}

\usepackage{graphicx,color}

\newcommand{\ket}[1]{\left | #1 \right \rangle}
\newcommand{\bra}[1]{\left \langle #1 \right |}

\def\openone{\leavevmode\hbox{\small1\kern-3.8pt\normalsize1}}

\def\cg{{\cal G}}
\def\ca{{\cal A}}

\def\ma{\mathbf{a}}
\def\mb{\mathbf{b}}
\def\mx{\mathbf{x}}
\def\mapr{\mathbf{a'}}

\def\ZZ{\mathbb{Z}}

\newtheorem{theorem}{Theorem}
\newtheorem{lemma}{Lemma}

\theoremstyle{definition}

\newcommand{\poly}{{\rm poly}}

\begin{document}
\title{\LARGE\bf
  Classical simulation complexity of\\
  extended Clifford circuits}
\author{
  Richard Jozsa$^1$
  and Maarten Van den Nest$^2$\\[3mm]
  \small\it
  \small\it $^1$DAMTP, Centre for Mathematical Sciences, University of Cambridge,\\ \small\it Wilberforce Road, Cambridge CB3 0WA, U.K.\\[1mm]
  \small\it $^2$Max Planck Institut f\"{u}r Quantenoptik, Hans-Kopfermann-Str. 1,\\ \small\it D-85748 Garching, Germany.\\[1mm]
}

\date{}

\maketitle

\begin{abstract}
 Clifford gates are a winsome class of quantum operations combining mathematical elegance with physical significance. The Gottesman-Knill theorem asserts that Clifford computations can be classically efficiently simulated but this is true only in a suitably restricted setting. Here we consider Clifford computations with a variety of additional ingredients: (a) strong vs. weak simulation, (b) inputs being computational basis states vs. general product states, (c) adaptive vs. non-adaptive choices of gates for circuits involving intermediate measurements, (d) single line outputs vs. multi-line outputs. We consider the classical simulation complexity of all combinations of these ingredients and show that many are not classically efficiently simulatable (subject to common complexity assumptions such as P not equal to NP). Our results reveal a surprising proximity of classical to quantum computing power viz. a class of classically simulatable quantum circuits which yields universal quantum computation if extended by a purely classical additional ingredient that does not extend the class of quantum processes occurring.
\end{abstract}

\section{Introduction}\label{intro}

\begin{figure}[ht]
\hspace{1.5cm}{\includegraphics[width=12cm]{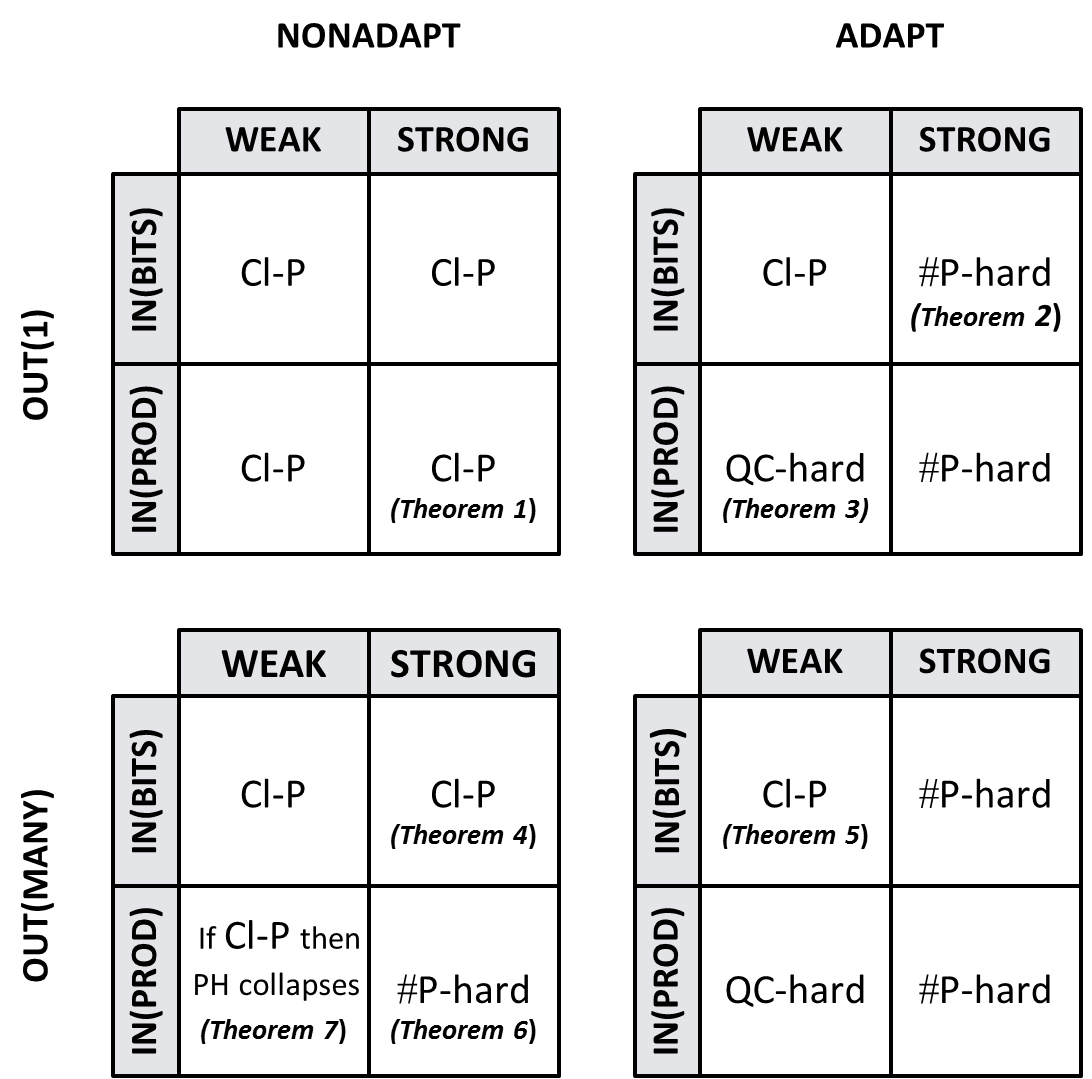}}
\caption[]{\label{Fig} Classical simulation complexities for sets of Clifford computational tasks. The acronyms are as defined in the main body of the text. The seven cases containing numbered theorems are proved in section \ref{proofs}. All other cases are easily seen to be implied by these seven cases.}
\end{figure}

The notion of classical simulation of quantum computation provides a mathematically precise tool for studying fundamental questions that are often only vaguely formulated -- questions of the relationship between classical and quantum computing power and the computational possibilities engendered by particular kinds of quantum resources. We may consider a restricted class $\ca$ of quantum circuits defined by specified limited quantum ingredients and ask whether it can be classically efficiently simulated or not. Computational hardness is notoriously difficult to establish and in the latter case we are generally content to establish that the efficient simulation of $\ca$ would imply some further property such as P$=$NP, that is widely regarded as implausible. In the former case of $\ca$ being classically efficiently simulatable we may consider enlarging $\ca$ to $\ca'$ by inclusion of some extra specific quantum ingredient $P$ and investigating the resulting change in classical simulation complexity. If for example, $\ca'$ allows universal quantum computation then in this mathematically precise sense, $P$ may be regarded as an ``essential resource for quantum computational power'' (relative to a background of quantum effects that are ``computationally lame''). Below we will see examples of seemingly quite modest expansions of resources leading to dramatic changes in simulation complexity, indicating that the computational landscape between classical and quantum computing power is a richly complex one.

This paper is devoted to developing a case study of simulation of Clifford circuits supplemented with a variety of extra ingredients. The choice of Clifford circuits is a particularly interesting and relevant one for a variety of reasons. Clifford computations  provide one of the earliest significant examples of classically simulatable quantum computations in the Gottesman-Knill theorem \cite{Go99} (see also \cite{nc, De03, Aa04, An06, Va10}), showing in particular that the presence of non-trivial entanglements in a quantum computation is not necessarily a signature of computational speed-up. Clifford gates also have a rich associated pure mathematical theory that may be drawn upon in the study of simulation properties, as well as having a rich physical and practical significance in the theory and implementation of quantum computation. For example Clifford operations feature prominently in the theory of quantum error correction and fault tolerance \cite{cliff, nc, Br05} and  in measurement-based quantum computation \cite{Ra01, Ra03}.

It is well known that the Clifford gates supplemented with any non-Clifford operation generate a dense subgroup of $U(2^n)$ and are hence universal for quantum computation \cite{Ne03, Ne06}. Here we will consider extensions of Clifford circuits of a different, perhaps seemingly more innocuous kind. More precisely we will characterise the classical simulation complexity of sixteen cases of extended Clifford circuits that are defined by four binary choices. Our main results are summarised in figure 1. The acronyms in figure 1 that define the extensions and their classical simulation complexities are all explained in detail in section \ref{prelim} below, and briefly they are as follows: IN(BITS) and IN(PROD) refer to allowing computational basis states and general product states as inputs. OUT(1) and OUT(MANY) refer to having single bit and multi-bit outputs. NONADAPT and ADAPT refer to circuits with intermediate measurements, with the circuit gates being respectively fixed or chosen adaptively as a function of previous measurement outcomes. WEAK and STRONG refer to two notions of classical simulation that provide respectively a sample of the output distribution and a calculation of actual probability values. In the body of the tables, Cl-P denotes that classical efficient simulation is possible, QC-hard denotes that universal quantum computation is possible, and \#P-hard asserts that classical simulation could be used to solve arbitrary problems in the classical class \#P (and hence NP too).

These results demonstrate a remarkable sensitivity of the classical simulation complexity of Clifford circuits under various small modifications. In section \ref{mainresults} we  highlight some interesting comparisons amongst these simulation complexities. In particular the issue of the last sentence of the abstract above is discussed in Example 2 of section \ref{mainresults}. Finally in section \ref{proofs} we provide proofs of all results given in figure 1. For completeness we indicate proofs for all sixteen cases. Some cases were previously known (cf references in our text) but to the best of our knowledge others have not previously been given in the literature.

Finally we mention here some related work on the classical simulation complexity of various extensions and generalizations of Clifford circuits. See \cite{Aa04} for simulation of  Clifford circuits supplemented with few non-stabilizer (pure or mixed) inputs and/or few non-Clifford gates; see \cite{Br05} for quantum computing with adaptive Clifford circuits with product state inputs (we will revisit this scenario as one of the sixteen cases in figure 1); see \cite{jm08, Va11} for simulations of Clifford circuits supplemented with certain non-Clifford gates by restricting the circuit structure; see \cite{Go99b, Ho05, De13, Va12, Be12} for generalizations of the Gottesman-Knill theorem to higher-dimensional systems; see \cite{cjl} for generalizations of Clifford circuits based on projective normalisers of finite unitary groups.

\section{Preliminary definitions and notations}\label{prelim}

\subsection*{Clifford circuits: NONADAPT and ADAPT}
Let $I,X,Y,Z$ denote the standard 1-qubit Pauli matrices \cite{nc} (amongst which we include the identity matrix). An {\em $n$-qubit Pauli operator} is any operator of the form $P=\gamma P_1\otimes \ldots \otimes P_n$ where $\gamma \in \{ \pm 1, \pm i \}$ and each $P_i$ is a Pauli matrix.

An $n$-qubit unitary operation $C$ is called a {\em Clifford operation} if the set of all Pauli operators is preserved under conjugation by $C$ i.e. for any
$n$-qubit Pauli operator $P$, $P'=CPC^\dagger$ is again a Pauli operator. It is known that $C$ is Clifford iff $C$ can be expressed as a circuit of the following gates (cf \cite{nc}):  the 1-qubit Hadamard gate $H$, the phase gate $T={\rm diag}(1, i)$ and the 2-qubit controlled-$Z$ gate $CZ$,
which we call {\em basic Clifford gates}. Moreover any $n$-qubit Clifford operation can be expressed as a circuit of $O(n^2)$ basic Clifford gates (see  \cite{De03} and theorem 10.6 of \cite{nc}).

A {\em unitary Clifford circuit} is a circuit comprising only the basic Clifford gates. The size of the circuit is the number of gates of which it consists.

As a further extension we will allow measurements in the body of the circuit. The term  {\em measurement} will always mean a single qubit measurement in the computational basis. Let $M_i(x)$ denote a measurement of the $i^{\rm th}$ qubit line with outcome $x\in \{ 0,1 \}$. Then a {\em Clifford circuit with $K$  intermediate measurements} has the form
\begin{equation}\label{cmmt} C_0\, M_{i_1}(x_1)\,C_1\, M_{i_2}(x_2)\, C_2\, \ldots \,M_{i_K}(x_{i_K})\,C_K \end{equation}
where $C_i$ are unitary Clifford circuits (possibly of size zero). We assume that measurements are non-destructive and the measured qubit, set to the designated post-measurement state, may generally be an input into subsequent operations e.g. $M_{i_1}(x_1)$ sets qubit line $i_1$ to $\ket{x_1}$ which may then be input into $C_1$.

A {\em non-adaptive Clifford circuit} is a Clifford circuit with intermediate measurements in which the choice of operations in the circuit does not depend on the outcomes of (previous) measurements. Hence such a circuit is fully defined by eq. (\ref{cmmt}) where the $C_j$'s and measurement line labels $i_j$'s are fixed a priori.

By the term {\em adaptive Clifford circuit} we will mean a process of the form eq. (\ref{cmmt})  in which the choice of operations is allowed to depend on previous measurement outcomes. To make this dependency explicit we can expand the notation of eq. (\ref{cmmt}) as
\begin{equation}\label{adaptcliff}
C_0\, M_{i_1}(x_1)\,C_1(x_1)\, M_{i_2(x_1)}(x_2)\, C_2(x_1,x_2) \ldots \,M_{i_K(x_1,\ldots x_{K-1})}(x_k)\,C_K(x_1, \ldots , x_K). \end{equation}
Note that the size of the circuits $C_j(x_1, \ldots , x_j)$ may vary with $x_1, \ldots , x_j$. The total number $N$ of operations in the adaptive circuit eq. (\ref{adaptcliff}) is defined to be the maximum number of elementary Clifford gates and measurements over all possible choices of measurement outcomes $x_1, \ldots , x_K$. Alternatively we could uniformise the size of each $C_j$ by including additional identity gates to make its size independent of $x_1, \ldots , x_j$. Similarly to uniformise the number of intermediate measurements in the adaptive process (as a function of measurement outcomes) we could formally allow $i_{j+1}(i_1, \ldots , i_j)$ to be zero (for qubit lines labelled 1 to $n$) to indicate that a measurement is not performed, but replaced by an identity gate.

The scenarios of non-adaptive and adaptive Clifford circuits will be denoted respectively by the acronyms NONADAPT and ADAPT.

We mention here two elementary simplifications of circuit structures that will be useful in proofs of classical simulation properties. Stated informally we have the following facts (with formal statements and proofs given in lemmas \ref{lemmacu} and \ref{lemmacad} in section \ref{proofs} below):\\
(i) without loss of generality (wlog) non-adaptive circuits may be assumed to be unitary;\\
(ii) in (adaptive or non-adaptive) circuits with intermediate measurements, wlog the measured qubits may be assumed to always be discarded after measurement (and not used in subsequent operations). Furthermore for adaptive circuits the choice of lines for intermediate measurements may be assumed to be non-adaptive.

\subsection*{Inputs and outputs: IN(BITS), IN(PROD), OUT(1) and OUT(MANY)}

In addition to the circuit itself there are two further ingredients for the full specification of a computational process viz. specification of the input and of the output. We will distinguish two classes of input states -- computational basis input states, denoted by the acronym IN(BITS), and general product state inputs, denoted IN(PROD). For outputs we will distinguish the scenarios of a single bit output (resulting from a specified final 1-qubit measurement), denoted OUT(1), and the scenario of a many-bit output, denoted OUT(MANY). In the latter case, for an $n$-qubit circuit the output $y_{j_1} \ldots y_{j_l}$ results from a final measurement on a specified set $ 1\leq j_1 < \ldots < j_l \leq n$ of $l$ lines and generally we can have $l=O(n)$.

\subsection*{Classical simulations: WEAK, STRONG and Cl-P}
A {\em description of a Clifford computational task $T$} with $N$ operations on $n$ qubits is made up of the following ingredients:\\
(i) a description of an (adaptive, non-adaptive or unitary) Clifford circuit on $n$ lines comprising $N$ operations. For unitary or non-adaptive circuits we give a list of $N$ basic Clifford gates and intermediate measurements on specified qubit lines; for adaptive circuits (cf eq. (\ref{adaptcliff})) we require that each $C_j(x_1, \ldots , x_j)$ and $i_{j+1}(i_1, \ldots , i_j)$ is given as a function computable in classical $\poly(N)$ time;\\ (ii) specification of an input state $\ket{\psi}$ which we always take to be either a computational basis state or a general product state;\\ (iii) specification of one or more output measurement lines $ 1\leq j_1 < \ldots < j_l \leq n$.\\ Let $p(y_{j_1}, \ldots , y_{j_l})$ denote the output probability distribution of the corresponding quantum process.

We will consider sets of computational tasks subject to the restrictions introduced above viz. the eight combinations of ADAPT vs. NONADAPT, IN(BITS) vs. IN(PROD) and OUT(1) vs. OUT(MANY). In each case it is natural to assume that the total length (as a classical bit string) of the full description (i), (ii) and (iii) of the computational task is $O(\poly(N))$. In particular we assume that there are no extraneous qubit lines that are not acted upon (so $n=O(N)$) and we assume that input product states are specified with $O(\poly(N))$ bits. The latter technical issue of accuracy (for our later purposes of simulation complexity characterisations)  may be addressed by setting up a suitable notion of approximation, but we do not elaborate it here for sake of clarity and conceptual transparency.

We introduce two notions of classical simulation for Clifford computational tasks. A {\em weak classical simulation} for a set of computational tasks is a classical randomised computation which, given a description of a task $T$ as input, outputs a {\em sample} of the output distribution $p(y_{j_1}, \ldots , y_{j_l})$ of  $T$. A {\em strong classical simulation} for a set of tasks is a classical computation whose input is a description of a task $T$ and bit values for a subset of its output lines. The output is the value of the corresponding marginal probability of the output distribution of $T$ i.e. we have a classical computation of any desired output probability or marginal probability of $T$.

A weak or strong classical simulation is called {\em efficient} if the corresponding classical computation runs in classical poly$(N)$ time. (Again here for strong simulation, as previously noted for IN(PROD), there is a further technical issue of precision and more formally we would require the output probability or marginal to be computed to $k$ bits of precision in poly$(N,k)$ time).

We will use the acronyms WEAK (resp. STRONG) to indicate that we are considering a weak (resp. strong) classical simulation for a set of tasks. We will use the acronym Cl-P (``classical poly time'') to assert that the associated simulation is efficient.

If the computational task $T$ is implemented as a quantum process it will require $O(N)$ quantum computational steps so existence of an efficient weak classical simulation implies that $T$ offers no quantum computational time benefit over classical computation (up to the usual polynomial overheads of resources commonly accepted in complexity theory).

In the case of OUT(MANY) there are generally exponentially  many output probabilities $p(y_{j_1}, \ldots , y_{j_l})$ (as $l$ may be $O(n)$) so in strong efficient simulation we cannot ask for a computation of them all. The inclusion of computation of marginals in the definition of strong simulation guarantees the following eponymously desirable result.

\begin{lemma}\label{lemmaterdiv} Let $p(x_1, \ldots , x_n)$ be a probability distribution over $n$ binary variables. Suppose that each of the $n$ marginals $p(x_1), p(x_1, x_2), \ldots , p(x_1, \ldots , x_n)$ may be efficiently classically computed for any choice of values $x_1, \ldots , x_n$, and suppose that any 1-bit distribution $\{ p_o, 1-p_o\}$ with $p_0$ efficiently computable, may be efficiently sampled. Then $p(x_1, \ldots , x_n)$ may be efficiently sampled.\\ Hence for any set of computational tasks, efficient strong classical simulation implies efficient weak classical simulation.
\end{lemma}
\noindent For a proof see proposition 1 of \cite{terdiv1}.

For completeness we mention that there are also notions of weak simulations which incorporate various types of approximations \cite{Va11, bjs}; these will however not be relevant for the present work.

\subsection*{Complexity measures: QC-hardness and \#P-hardness}

We will also be interested in establishing that some sets of Clifford computational tasks are unlikely to have efficient classical simulations and for this purpose we introduce some further complexity notions.

Consider the following classical computational task called \#SAT (cf \cite{AB}):\\ {\sc Input}: a Boolean function $f$ from $n$ bits to one bit (given say as a bit string encoding a formula in standard 3-cnf form \cite{AB}).\\ {\sc Problem:} determine the number \#$f$ of $n$-bit strings $\mx$ with $f(\mx )=1$.\\ Note that \#SAT is a generalisation of the well known NP-complete problem SAT \cite{AB} (which asks only if \#$f$ is non-zero) so it is very unlikely that \#SAT has a poly time classical algorithm; indeed the latter would imply equality of the complexity classes P and NP, and also that any problem in \#P may be computed in poly time \cite{AB}.

A set $\ca$ of Clifford computational tasks $T$ is called {\em \#P-hard} if an efficient strong simulation of $\ca$ would give rise to an efficient classical solution of \#SAT. More precisely $\ca$ is \#P-hard if given any input $f$ of size $N$ for the \#SAT problem, it may be converted by a classical poly$(N)$ time computation $\phi$ into (a description of) a task $\phi(f)$ in $\ca$ with the following property: \#$f$ may be computed by a classical poly$(N)$ time algorithm  from the results of strong classical simulation of  $\phi(f)$. Hence efficient strong classical simulation of $\ca$ would imply P$=$NP and that \#P is computable in poly time.

Finally we introduce a notion of {\em QC-hardness} for a set $\ca$ of Clifford computational tasks. This will be used to indicate that $\ca$ is unlikely to have an efficient weak classical simulation. Broadly speaking $\ca$ will be QC-hard (``quantum computing hard'') if it is rich enough to encode universal quantum computation, so then efficient classical weak simulation of $\ca$ would imply that all quantum computation could be classically efficiently simulated. More precisely we will adopt the following definition. Let $\cg$ be the set of basic unitary Clifford gates together with the phase gate
$S= \mbox{diag}(1, e^{i\pi/4})$. It is known \cite{nc} that $\cg$ is a universal set of gates for quantum computation. Let $C$ be any circuit of gates from $\cg$ with a specified computational basis state input and 1-bit output from measurement of a specified output line. Then $\ca$ is QC-hard if any such $C$ may be simulated by a member $\phi(C)$ of $\ca$ i.e. $T=\phi(C)$ has 1-bit output whose probability distribution for the given computational basis input coincides with that of $C$. Here as before, $\phi$ is a poly time translation of the description of $C$ into the description of a member of $\ca$. Hence efficient weak classical simulation of $\ca$ would imply efficient classical simulation of universal quantum computing.

\section{Main results -- statement and discussion}\label{mainresults}

We now consider the sixteen sets of Clifford computational tasks defined by all combinations of the following four binary choices

\vspace{2mm}

(i) NONADAPT vs. ADAPT

(ii) IN(BITS) vs. IN(PROD)

(iii) OUT(1) vs. OUT(MANY)

(iv) WEAK vs. STRONG.

\vspace{2mm}

\noindent The corresponding simulation complexities that we will prove, are summarised in figure 1.

The original Gottesman-Knill  theorem \cite{Go99} asserts that efficient classical simulations exist for the case ADAPT, IN(BITS), OUT(MANY) and WEAK (cf. theorem \ref{thm5} below). In contrast, we find here that eight of the sixteen cases of extended Clifford circuits are not (likely to be) classically efficiently simulatable.

We draw attention to some examples of extreme changes of simulation complexity resulting from seemingly modest modifications in the defining computational resources. Perhaps the most significant such comparisons for issues of computing power will be cases involving only weak simulations, since implementing the circuit itself as a quantum process yields only one sample of the output probability distribution, in contrast to the far greater information resulting from strong simulation (cf. \cite{Va10, Va11}).

\noindent {\bf Example 1.}  For the case of non-adaptive (or equivalently, unitary) circuits with general product state inputs, we have that 1-bit outputs are classically efficiently simulatable in both weak and strong senses. However allowing just many bit outputs results in \#P hardness for strong simulation and a more subtle certification of hardness (related to PH collapse) for weak simulation. This indicates a significant increase of computational power associated to the passage from OUT(1) to OUT(MANY) i.e. merely sampling more lines of the same class of quantum processes (and see also \cite{bjs} where  a similar phenomenon is observed for computational processes defined by commuting quantum circuits).

\noindent On the other hand, in the case of adaptive circuits with computational basis inputs the passage from one to many output lines remains classically efficiently simulatable in the weak sense (all other adaptive cases already being QC- or \#P-hard).

\noindent Note also that in our definitions of classical simulation we ask for simulation only of the output distribution of the computational task, and not of intermediate measurement distributions (if there are any).
Indeed inclusion of the latter could elevate an OUT(1) scenario to
OUT(MANY) via consideration of intermediate measurement outcomes together with the single bit output of the task, and the associated simulation complexity could radically change. $\Box$

\noindent {\bf Example 2.} Another particularly interesting comparison is that of weak simulation for general product state inputs and single bit outputs, with the transition from non-adaptive to adaptive circuits i.e. we compare\\ Case A: IN(PROD), OUT(1), WEAK,  NONADAPT to\\ Case B: IN(PROD), OUT(1), WEAK,  ADAPT.\\  Case A admits efficient weak classical simulation whereas case B is QC-hard.
But now note that the passage from Case A to Case B involves the inclusion of a purely {\em classical} extra resource viz. classical adaptive choice of gates, without introducing any new gates. Furthermore the class of quantum processes occurring in runs of Case B is {\em exactly the same} as the class occurring in runs of Case A, since any single actual run of an adaptive circuit can occur as a run of a non-adaptive circuit (that non-adaptively prescribes the sequence of gates that were adaptively chosen). Indeed from an experimentalist's point of view cases A and B may be claimed to be totally indistinguishable in the following sense: suppose an experimentalist $\cal E$ has the ability to implement basic Clifford gates and measurements. A theorist $\cal T$ directs $\cal E$ by announcing one by one, a sequence of basic gates and measurements, which $\cal E$ successively implements. For each measurement instruction $\cal E$ announces the measurement outcome before further instructions from $\cal T$. Then $\cal E$ cannot tell whether $\cal T$ is choosing gates adaptively (Case B) or not (Case A) -- the demands on $\cal E$'s laboratory are exactly the same, and case B results in no new quantum processes. Yet Case B can perform universal quantum computation whereas Case A is fully classically efficiently simulatable. $\Box$

The sixteen cases of extended Clifford circuits give rise to a rich landscape of simulation complexities. Apart from Cl-P, QC-hardness and \#P-hardness,  we will see in the course of the proofs that connections to other major complexity classes appear as well. For example (cf. remark below theorem \ref{thm5}),  uniform families of adaptive Clifford circuits with computational basis inputs have precisely the same power as a universal randomised classical computation. Thus the class of languages decidable by such Clifford processes in poly-time with bounded error, is precisely BPP. What is more, just changing from computational basis inputs to arbitrary product state inputs (and keeping the other parameters equal) yields universal quantum computation, so in the same poly-time bounded error setting, these Clifford processes now give precisely  BQP \cite{Br05}. We will also find that post-selected non-adaptive Clifford circuits with product state inputs have the same power as BQP with postselection (cf. theorem \ref{thm7} and \cite{Br05}), which is known to coincide with the class PP \cite{aarpp}. Finally we recall that the simulation complexity of non-adaptive Clifford circuits with computational basis inputs and single bit outputs is known to be characterized  by the class $\oplus$L $\subseteq $ P \cite{Aa04}.

\section{Proofs of main results}\label{proofs}

In this section we give proofs of theorems 1 to 7 that appear as seven of the sixteen cases depicted in figure 1. For the remaining cases it may be easily checked that they all follow from the seven basic cases using the following simple facts:\\
(i) if  a set of tasks is Cl-P then any subset is Cl-P too;\\
(ii) if a subset of tasks is QC- or \#P-hard then the full set is QC- or \#P-hard too;\\
(iii) replacing IN(BITS) by IN(PROD), or replacing OUT(1) by OUT(MANY), increases the set of computational tasks;\\
(iv) by lemma \ref{lemmaterdiv}, if strong simulation is Cl-P, then weak simulation is Cl-P too (keeping all other resource choices unchanged).

We will use the following notations relating to bit strings and Pauli operators.
 For any $n$-bit strings $\mathbf{a}=a_1\ldots a_n$ and $\mathbf{b}=b_1\ldots b_n$, $\mathbf{c}= \mathbf{a}+\mathbf{b}$ will denote the $n$-bit string with $c_i=a_i\oplus b_i$ (and $\oplus$ being addition mod 2), and $\mathbf{a}\cdot \mathbf{b}=a_1b_1\oplus \ldots \oplus a_nb_n$ will denote the mod 2 inner product. We will also use the notation $X(\mathbf{a})=X^{a_1}\otimes \ldots \otimes X^{a_n}$ and $Z(\mathbf{a})=Z^{a_1}\otimes \ldots \otimes Z^{a_n}$. $\ket{\mathbf{a}}$ will denote the computational basis state corresponding to $\mathbf{a}$.
Then the following properties are easily verified for any $n$-bit strings $\mathbf{x,a,a'}$:
\begin{equation}\label{startpauli}\begin{array}{l}
X(\mathbf{a})\ket{\mathbf{x}}=\ket{\mathbf{x}+\mathbf{a}), \hspace{5mm} Z(\ma}\ket{\mx}=(-1)^{\mx\cdot \ma}\ket{\mx}, \\
 X(\ma)X(\mapr)=X(\ma+\mapr), \hspace{5mm}Z(\ma)Z(\mapr)=Z(\ma+\mapr) \\
X(\ma)Z(\mapr)=(-1)^{\ma\cdot \mapr}Z(\mapr)X(\ma). \end{array}
\end{equation}
Since $Y=iXZ$, any Pauli operator $P$ can be written uniquely as $P=\alpha X(\ma)Z(\mb)$ for some $\alpha \in \{ \pm 1, \pm i \}$ and $n$-bit strings $\ma$ and $\mb$. Labelling the $\alpha$ values by 2-bit strings $r_1r_2$, we call the $(2n+2)$-bit string $(r_1r_2, \ma ,\mb )$ the {\em label} of $P$ and $\alpha$ the {\em phase} of $P$.
 If $G$ is any basic Clifford gate (acting on specified qubit line(s), and extended by $I$ on all other lines), and $P$ is any $n$-qubit Pauli operator, then the label of $P'=GPG^\dagger$ can be easily computed from the label of $P$ in $O(n)$ time. In fact only the phase of $P$ and the label entries pertaining to the line(s) of action of the basic Clifford gate are modified.

We begin by proving two elementary simplifications of circuit structures that were mentioned in section \ref{prelim}.
To formally establish these we use the following construction: let $C$ be any (adaptive or non-adaptive) circuit on $n$ lines with $K$ intermediate measurements, input state $\ket{\psi}$ and final output measurements on lines $ 1\leq j_1  \ldots < j_l \leq n$.  Introduce an enlarged unitary circuit $C^*$ on $n+K$ lines defined as follows. For each intermediate measurement $M_{i}$ on line $i$ of $C$ introduce an extra ancilla qubit (line $n+i$) in state $\ket{0}$ and replace the measurement operation by the unitary Clifford operation $CX_{i,n+i}$ (where $CX_{j,k}$ is the 2-qubit controlled-$X$ operation with source $j$ and target $k$).

\begin{lemma} \label{lemmacu} Suppose $C$ with input and output as above is a non-adaptive circuit with $K$ intermediate measurements. Then there is a {\em unitary} circuit $C'$ on $n+K$ lines which is equivalent to $C$ in the following sense: if $C'$ has input $\ket{\psi}\ket{0}\ldots \ket{0}$ then measurement of lines $j_1,\ldots j_l$ of $C'$ will result in the same probability distribution of outputs as $C$ on $\ket{\psi}$.
\end{lemma}
\noindent{\bf Proof.} We just take $C'$ to be $C^*$ as defined above. $\Box$

\begin{lemma}\label{lemmacad}  Suppose $C$ as above is an adaptive circuit. Then there is an adaptive circuit $\tilde{C}$ on $n+K$ lines which is equivalent to $C$ (in the sense given in lemma \ref{lemmacu} above) and\\   (i) in $\tilde{C}$ after each intermediate measurement $M_i(x_i)$ the line $i$ and its post-measurement state are not further used in any subsequent operations of $\tilde{C}$. Furthermore the choice of line $i$ here is always non-adaptive i.e. independent of previous measurement outcomes.\\
(ii) In $\tilde{C}$ the set of output lines $\{ j_1, \ldots , j_l \}$ is disjoint from the set of intermediate measured lines.
\end{lemma}
\noindent{\bf Proof.} To construct $\tilde{C}$ we take $C^*$ as above, but after each extra $CX_{i,n+i}$ operation we immediately measure line $n+i$, and use its output as the result of the intermediate measurement $M_i$ of $C$, for subsequent adaptations. $\Box$

\noindent We are now ready to prove our seven theorems.

\subsection*{NONADAPT, IN(PROD),  OUT(1) and STRONG: Cl-P}

\begin{theorem}\label{thm1} Let $\ca$ be the set of computational tasks defined by non-adaptive Clifford circuits, general product state inputs and single bit outputs. Then $\ca$ may be strongly efficiently classically simulated.
\end{theorem}

\noindent {\bf Proof.} This result has been proved in \cite{cjl} and we summarise the argument here. Using lemma \ref{lemmacu} we may assume wlog that the Clifford circuit is unitary. Let $C=C_N\ldots C_1$ be a unitary Clifford circuit with product state input $\ket{\alpha}=\ket{\alpha_1}\ldots \ket{\alpha_n}$. Write $\ket{\beta}=C\ket{\alpha}$. We may assume that the output, with probabilities $p_0,p_1$  is obtained from line 1 (as the swap gate is Clifford). Let $A=Z\otimes I\otimes I\ldots \otimes I$. Then $p_0-p_1= \bra{\beta}A\ket{\beta}=\bra{\alpha}C^\dagger A C\ket{\alpha}$. Now $A$ is a Pauli operator so after successive conjugations by the $C_i$'s we get $C^\dagger AC=\gamma P_1\otimes \ldots \otimes P_n$ where the label of the latter is easily computed in $\poly (N)$ time. Thus $p_0-p_1= \gamma \prod_{i=1}^n \bra{\alpha_i}P_i\ket{\alpha_i}$ and the latter expression, being a product of $n$ $2 \times 2$ matrix expectation values, is readily computable in $\poly (N)$ time, providing the efficient strong classical simulation (as $p_0+p_1=1$ too). $\Box$

\noindent {\bf Remark.} The simple method of the above proof does not generalise to the case of OUT(MANY) with $O(n)$ output lines. Indeed we will see (cf theorems \ref{thm4} and \ref{thm6} below) that this case is \#P-hard but remains classically efficiently strongly simulatable if we restrict the input states to just computational basis states i.e. to IN(BITS).

\subsection*{ADAPT, IN(BITS),  OUT(1) and STRONG: \#P-hard}

\begin{theorem}\label{thm2} Let $\ca$ be the set of computational tasks defined by adaptive Clifford circuits, computational basis state inputs and single bit outputs. Then the strong classical simulation of $\ca$ is \#P-hard.
\end{theorem}

\noindent {\bf Proof.} With the availability of adaptation we are able to apply the gate $CX_{jk}$ or the identity gate $I_{jk}$ (on lines $j$ and $k$) chosen conditionally according to the result of a measurement on another line $i$. Thus if these lines are promised to be in computational basis states we can apply the Toffoli gate. (Note however that we cannot by this method apply the Toffoli gate coherently on general quantum states because the adaptation requires a measurement on line $i$). Then with the availability of computational basis state inputs, using $X$ and this Toffoli construction, we can efficiently implement universal classical computation. Thus if $f$ is any Boolean function from $n$ bits to one bit, we can implement the transformation $A_f: \ket{\mx}\ket{0} \rightarrow \ket{\mx}\ket{f(\mx )}$ (so long as the input is a computational  basis state).  Consider now the following process which is allowed in $\ca$: starting with $n$ qubits each  in state $\ket{0}$, apply $H$ to each and measure each to generate a uniformly random $n$-qubit computational basis state $\ket{\mx}$. Then apply $A_f$ and finally measure the qubit line of $\ket{f(\mx )}$ to give a single bit output. Clearly the probability of obtaining 1 is $\#f/2^n$ so strong simulation of this process is \#P-hard. $\Box$

\subsection*{ADAPT, IN(PROD),  OUT(1) and WEAK: QC-hard}

\begin{theorem}\label{thm3} Let $\ca$ be the set of computational tasks defined by adaptive Clifford circuits, general product state inputs and single bit outputs. Then the weak classical simulation of $\ca$ is QC-hard.
\end{theorem}

\noindent {\bf Proof.} This result is well known, see e.g. \cite{Br05}. It suffices to show that within the given resource constraints, the phase gate $S={\rm diag}(1,e^{i\pi/4})$ may be implemented on any desired qubit line. This is achieved by introducing an extra ancilla qubit labelled $a$,  in state $\ket{\frac{\pi}{4}}=\frac{1}{\sqrt{2}}(\ket{0}+e^{i\pi/4}\ket{1})$ (respecting availability of product state inputs) and then applying the process of lemma \ref{sgate} below, and finally applying the Clifford gate $T=S^2$ to line $i$ conditionally on the value of the ancilla measurement outcome (which is possible since adaptation is available). $\Box$

\begin{lemma}\label{sgate} Let $\ket{\psi}_{1\ldots n}$ be an $n$-qubit state on lines 1 to $n$ and let $S={\rm diag}(1,e^{i\pi/4})$. Let  $\ket{\frac{\pi}{4}}_a=\frac{1}{\sqrt{2}}(\ket{0}+e^{i\pi/4}\ket{1})$ be an extra ancillary qubit. Then
 \[  \mbox{ $M_a(x)\, CX_{ai}\, \ket{\psi}\ket{\frac{\pi}{4}}_a$ results in}\hspace{2mm} \left\{ \begin{array}{rl}
S_i\ket{\psi}\ket{0}_a & \mbox{if $x=0$} \\
e^{i\pi/4}S_i^{-1}\ket{\psi}\ket{1}_a & \mbox{if $x=1$}
\end{array}\right. \]
where $S_i$ denotes the application of $S$ to qubit $i$, and $CX_{ai}$ is the application of $CX$ to lines $a$ and $i$ with $i$ as target line.
\end{lemma}

\noindent {\bf Proof of lemma.} A straightforward calculation. $\Box$

\subsection*{NONADAPT, IN(BITS),  OUT(MANY) and STRONG: Cl-P}

\begin{theorem}\label{thm4} Let $\ca$ be the set of computational tasks defined by non-adaptive Clifford circuits, computational basis state inputs and multiple bit outputs. Then $\ca$ may be strongly efficiently classically simulated.
\end{theorem}

\noindent {\bf Proof.} The techniques of \cite{De03} and alternatively \cite{Va10} may be used to prove theorem \ref{thm4}. Here we give a proof using a third method. Let $C$ be a non-adaptive Clifford circuit with computational basis input $\ket{x_1\ldots x_n}$ and let $j_1, \ldots ,j_m$ be any subset of the output lines. We will show that the corresponding marginal probability $p(y_1, \ldots , y_m)$ may be efficiently classically computed. We may assume the following standardised situation:\\
(i) $C$ is unitary (by lemma \ref{lemmacu});\\
(ii) $x_1\ldots x_n = 00\ldots 0$ and $y_1\ldots y_m= 00\ldots 0$ (since we can pre- and post- include extra $X$ gates on lines where $x_i$ or $y_j$ are 1);\\
(iii) $j_1, \ldots , j_m= 1,\ldots ,m$ for $m\leq n$ (since swap gates are Clifford operations).\\
Thus for $C$ unitary with input $\ket{0^n}=\ket{00\ldots 0}$ let $p={\rm prob}(0\ldots 0)$ be the probability of obtaining 0 from measurement of each of the lines 1 to $m$. Using $\ket{0}\bra{0}=(I+Z)/2$ and writing $\mathbf{t}=t_1\ldots t_m$ for $m$-bit strings we have
\[ \begin{array}{rcl}p&=&\frac{1}{2^m}\bra{0^n}C^\dagger \,\,(I_1+Z_1)\otimes \ldots \otimes (I_m+Z_m) \otimes I_{m+1}\otimes\ldots \otimes I_{n}\,\,C \ket{0^n} \\
 &=& \frac{1}{2^m} \sum_{\mathbf{t}\in \ZZ_2^m} \bra{0^n}C^\dagger\,\, \tilde{Z}(\mathbf{t})\,\,C\ket{0^n} \end{array}
\]
where $\tilde{Z}(\mathbf{t})$ is the $n$-qubit Pauli operator $Z(\mathbf{t})\otimes I \ldots \otimes I$ obtained by extending the $m$-qubit operator $Z(\mathbf{t})$ by $(n-m)$ $I$'s. This is a sum with potentially exponentially many terms (e.g. if $m=O(n)$) yet it can be evaluated in poly$(n)$ time as follows. Using the Clifford conjugation relations we have
\begin{equation}\label{cliff}
\Gamma (\mathbf{t}) \equiv  C^\dagger\, \tilde{Z}(\mathbf{t})\,C = \sigma(\mathbf{t})X(\mathbf{a}(\mathbf{t}))Z(\mathbf{b}(\mathbf{t}))
\end{equation}
with $\sigma(\mathbf{t})\in \{ \pm 1,\pm i \}$ and $\mathbf{a}(\mathbf{t}), \mathbf{b}(\mathbf{t}) \in \ZZ_2^n$.
Furthermore,  for each $\mathbf{t}$ these labels can be computed efficiently.

Next, introduce basis vectors $\mathbf{e}_j = 0\ldots 010\ldots 0$ in $\ZZ_2^m$ (having 1 in the $j^{\rm th}$ slot) for $j=1,\ldots ,m$. Then since $\mathbf{t}=\sum_i t_i \mathbf{e}_i$  and $\tilde{Z}(\mathbf{t})=\tilde{Z}(\mathbf{e}_1)^{t_1} \ldots \tilde{Z}(\mathbf{e}_m)^{t_m}$ we have
\begin{equation} \label{basis} \ma (\mathbf{t})=\sum_{i=1}^m t_i \,\ma (\mathbf{e}_i). \end{equation}
Next note that since $\bra{0}X\ket{0}=0$ we have
\[ \bra{0^n}\Gamma (\mathbf{t}) \ket{0^n} \neq 0 \hspace{3mm} \mbox{iff} \hspace{3mm} \ma (\mathbf{t}) = 0^n .\]
Furthermore if $\ma(\mathbf{t}) = 0^n$ then \begin{equation}\label{star1} \Gamma (\mathbf{t})=\sigma (\mathbf{t})\,Z(\mb (\mathbf{t})) \end{equation}
and since $\Gamma (\mathbf{t})^2=I = Z(\mb (\mathbf{t}))^2$ we must have $\sigma (\mathbf{t}) \in \{ \pm 1\}$, so that \begin{equation}\label{star2} \sigma (\mathbf{t})= (-1)^{u(\mathbf{t})} \end{equation}
with $u(\mathbf{t}) \in\{0, 1\} $. Furthermore, using $\bra{0}Z\ket{0}=\bra{0}I\ket{0}=1$ we get
\[ \bra{0^n}\Gamma (\mathbf{t}) \ket{0^n} = \sigma (\mathbf{t}) \hspace{3mm} \mbox{if} \hspace{3mm} \ma (\mathbf{t}) = 0^n .\]
Introducing $T_0=\{ \mathbf{t}:\ma (\mathbf{t})=0^n \} \subseteq \ZZ_2^m$
we thus get
\[ p= \frac{1}{2^m} \sum_{\mathbf{t}\in T_0}  (-1)^{u(\mathbf{t})}. \]
Next we characterise $T_0$. We have $\mathbf{t} \in T_0$ iff $\ma(\mathbf{t})=0^n$ so by eq. (\ref{basis}), $T_0$ is the subspace of $\ZZ_2^m$ given by the solution space of $A\mathbf{t}=0$ where $A$ is the $n\times m$ sized matrix with $\ma(\mathbf{e}_i)$ for $i=1, \ldots m$ as the columns. Using the label update rules for Clifford conjugations, all  $\ma(\mathbf{e}_i)$'s can be computed in $\poly (n)$ time. Thus we can compute a basis $\{ \mathbf{c}_1, \ldots , \mathbf{c}_l \}$ of $T_0$ (and hence also the information of  its dimension $l$) in $\poly(n)$ time. Then $\mathbf{t} \in T_0$ iff $\mathbf{t}=\sum_{i=1}^l s_i\mathbf{c}_i$ for $\mathbf{s}=s_1\ldots s_l\in \ZZ_2^l$ and
\[ p= \frac{1}{2^m} \sum_{\mathbf{s}\in \ZZ_2^l} (-1)^{u(\sum s_i\mathbf{c}_i)}. \]
Finally recalling that $Z(\mathbf{t}+\mathbf{t'})=Z(\mathbf{t})Z(\mathbf{t'})$ we see from eqs (\ref{star1}) and (\ref{star2}) that $u(\mathbf{t})$ is a linear function of $\mathbf{t}$, so writing $u(\mathbf{c}_i)=k_i$ and $\mathbf{k}=k_1\ldots k_l$ we have
\[ p= \frac{1}{2^m} \sum_{\mathbf{s}\in \ZZ_2^l} (-1)^{\mathbf{k}\cdot \mathbf{s}}. \]
Now $g(\mathbf{s})=(-1)^{\mathbf{k}\cdot \mathbf{s}}$ is a balanced function for $\mathbf{k}\neq 0^l$ (i.e. taking values $\pm 1$ equally often) so
\[ p= \left\{ \begin{array}{cl} 2^l/2^m & \rm{ if} \hspace{3mm} \mbox{$\mathbf{k}= 0^l$}\\
0 & \rm{ if} \hspace{3mm}\mbox{$\mathbf{k}\neq 0^l$} \end{array} \right. \]
concluding our efficient classical computation of $p$.\\
To summarise: given the description of the circuit $C$ we first compute $\ma(\mathbf{e}_i)$ for $i=1,\ldots m$ (from the Clifford conjugation relations) giving the matrix $A$ via columns. Then we compute any basis $\{ \mathbf{c}_1, \ldots , \mathbf{c}_l \}$ of ker$(A)$, and compute the $l$-bit string $\mathbf{k}=
u(\mathbf{c}_1)\ldots u(\mathbf{c}_l)$ (again from the Clifford conjugation relations in eq. (\ref{cliff}) with $\mathbf{t}=\mathbf{c}_i$ there). Then $p=2^{l-m}$ if $\mathbf{k}=0^l$ and $p=0$ otherwise. $\Box$

\subsection*{ADAPT, IN(BITS),  OUT(MANY) and WEAK: Cl-P}

\begin{theorem}\label{thm5} Let $\ca$ be the set of computational tasks defined by adaptive Clifford circuits, computational basis state inputs and multiple bit outputs. Then $\ca$ may be weakly efficiently classically simulated.
\end{theorem}

\noindent {\bf Remark.} Note that by theorem \ref{thm2} strong simulation in this scenario even with {\em single} bit outputs, is \#P-hard.
The weak simulation that we give in the proof of theorem \ref{thm5} below will use the strong simulation result of theorem \ref{thm4}. A different proof of theorem \ref{thm5} may be given in terms of the stabiliser formalism (see \cite{nc}, especially the Gottesman Knill theorem 10.7 therein) which develops a description of the evolving state through the course of the computation.

\noindent {\bf Remark.} A family of Clifford computational tasks $\{T_n: n=1, 2, \dots\}$, where $T_n$ acts on $n$ qubits, is said to be uniform if the description of $T_n$ can be computed in poly$(n)$ time by a (deterministic) classical Turing machine on input of $n$. Theorem \ref{thm5} shows that uniform families of adaptive Clifford circuits with computational basis state inputs and multiple bit outputs do not have additional power over polynomial-time randomised classical computation. Interestingly, the power of such uniform families of Clifford computational tasks in fact \emph{coincides} with  polynomial-time randomised classical computation. This follows from the constructions in the proof of theorem \ref{thm2} where it was shown how to generate  random bits and realize Toffoli gates with adaptive Clifford circuits acting on computational basis state inputs (see also \cite{An09} for related insights on realizing universal classical computation with adaptive stabilizer measurements). Finally we note the interesting comparison with the case of product states (replacing computational basis states) as inputs (keeping all other parameters the same) where the associated uniform families of Clifford computational tasks have precisely the same power as universal \emph{quantum} computation (which similarly immediately follows from the proof of theorem \ref{thm3}).

\noindent {\bf Proof.}  Let $C$ be an adaptive circuit on $n$ qubit lines with $K$ intermediate measurements, input $\ket{\mx}=\ket{x_1}\ldots \ket{x_n}$ and $l$ output lines. By lemma \ref{lemmacad} we may wlog instead work with an extended circuit $\tilde{C}$ on $n+K$ lines having the following form (rearranging the order of lines in lemma \ref{lemmacad}): $\tilde{C}$ has input $\ket{0}_1\ldots \ket{0}_K\ket{x_1}_{K+1}\ldots \ket{x_n}_{K+n}$ and the output measurements are on lines $K+1, \ldots , K+l$ (wlog, as swap operations are Clifford). Furthermore the $i^{\rm th}$ intermediate measurement yielding outcome $y_i$ for $1\leq i\leq K$ is on line $i$ and then line $i$ is not further used in $\tilde{C}$. As such, these measurements can be viewed as outputs too with the caveat that subsequent choices of gates may depend on the values $y_1,y_2, \ldots , y_K$ as they sequentially emerge. A full run of $\tilde{C}$ (including its $l$ output measurements) samples an associated probability distribution $p(y_1, \ldots , y_K,y_{K+1}, \ldots , y_{K+l})$.

Now if $y_1, \ldots , y_j$ for $j\leq K$ are specified then the circuit up to the $j^{\rm th}$ measurement becomes non-adaptive (i.e. the adaptive choices have been specified) and hence we can efficiently compute the marginal $p(y_1, \ldots , y_j)$ by theorem \ref{thm4}. Similarly if $j\geq K+1$ all adaptations have been specified and by theorem \ref{thm4} we can again efficiently compute the corresponding marginals $p(y_1, \ldots , y_j)$. Hence by lemma \ref{lemmaterdiv} we can efficiently sample the distribution $p(y_1, \ldots , y_K,y_{K+1}, \ldots , y_{K+l})$ and the last $l$ bits of the sample provides a weak efficient classical simulation of $\tilde{C}$ and hence of $C$ too. $\Box$

\subsection*{NONADAPT, IN(PROD),  OUT(MANY) and STRONG: \#P-hard}

\begin{theorem}\label{thm6} Let $\ca$ be the set of computational tasks defined by non-adaptive Clifford circuits, general product state inputs and multiple bit outputs. Then the strong classical simulation of $\ca$ is \#P-hard.
\end{theorem}

\noindent {\bf Remark.} Note that by theorem \ref{thm1} the same scenario with just 1-bit outputs is classically strongly efficiently simulatable.

\noindent {\bf Proof.}  We will show that efficient strong simulation of $\ca$ would imply efficient strong simulation of universal quantum computation and hence provide an efficient solution of the \#SAT problem (using the process described in the proof of theorem \ref{thm2} to express $\#f$ for any Boolean $f$ as a probability value).

Thus let $D$ be any quantum circuit comprising basic Clifford gates and $S$ gates with a product state input, and single bit output denoted $y$. Consider again the process of lemma \ref{sgate}. In our present scenario for $\ca$ we do not have adaptation available so we cannot implement $S$ gates as we did in the proof of theorem \ref{thm3}. Instead we proceed as follows. Suppose there are $K$ S gates in $D$. For each such gate introduce an ancilla in state $\ket{\frac{\pi}{4}}$ and replace the $S$ gate by the sequence of operations in lemma \ref{sgate}, resulting in a non-adaptive circuit $D'$ now involving only basic Clifford gates. Then $D'$ has $K+1$ outputs viz. $y$ and measurements of the $K$ ancilla lines denoted $a_1, \ldots , a_K$, and we have
\[ \begin{array}{rcl} {\rm Prob}_D(y)& = & {\rm Prob}_{D'}(y\, |\, 0_{a_1}\ldots 0_{a_K}) \\
 &=& {\rm Prob}_{D'}(y \, 0_{a_1}\ldots 0_{a_K})/{\rm Prob}_{D'}(0_{a_1}\ldots 0_{a_K}). \end{array} \]
Strong classical efficient simulation of $\ca$ (which allows multi-line outputs) implies that we can compute both of the $D'$ probabilities in the above quotient and hence ${\rm Prob}_D(y)$ i.e. we then get a strong efficient simulation of $D$.
 $\Box$

\subsection*{NONADAPT, IN(PROD),  OUT(MANY) and WEAK: collapse of PH}

\begin{theorem}\label{thm7} Let $\ca$ be the set of computational tasks defined (as in theorem \ref{thm6}) by non-adaptive Clifford circuits, general product state inputs and multiple bit outputs. If  $\ca$ could be weakly efficiently classically simulated, then the polynomial hierarchy PH would collapse to its third level.
\end{theorem}

\noindent {\bf Remark.} For the definition of PH we refer to \cite{AB}. The proof of theorem \ref{thm7} below rests on techniques introduced in \cite{bjs} and below we will be content to describe the relationship of the class $\ca$ in theorem \ref{thm7} to the constructions of \cite{bjs} and refer to the latter for further details of the proof.

\noindent {\bf Remark.} Theorem \ref{thm7} provides a partial answer to an open problem raised in \cite{Aa04} viz. the question of  the computational power of non-adaptive Clifford circuits with product state inputs and multiple bit outputs.

\noindent {\bf Proof.} Consider again the process of lemma \ref{sgate}. Now instead of utilising adaptation (as we did in the proof of theorem \ref{thm3}) or conditional probabilities (as we did above in theorem \ref{thm6}), we could alternatively implement $S$ using the process of lemma \ref{sgate} if we were able to {\em post-select} on measurement outcomes viz. we post-select the value 0 of the ancilla measurement. It follows that our class $\ca$ {\em together with post-selection} contains universal quantum computation, and even more, universal quantum computation with post-selection. Aaronson \cite{aarpp} has shown that the class BQP with post-selection coincides with the classical class PP (cf \cite{AB} for definitions). Thus our class $\ca$ with post-selection contains PP.

Now let $\cal K$ be any class of bounded error quantum circuits such that $\cal K$ with post-selection contains PP. Then (as elaborated in \cite{bjs}) weak efficient classical simulation of $\cal K$ for output measurements on many lines, implies that $\cal K$ with post-selection is contained in BPP with post-selection \cite{bjs}. Then according to  a result of classical complexity theory (cf \cite{bjs} for details), the latter inclusion (implying that PP is contained in BPP with post-selection) implies that PH collapses to its third level. Hence weak efficient classical simulation of our class $\ca$ would imply this collapse. $\Box$

\subsection*{Acknowledgments}
RJ was supported in part by the EC networks Q-ESSENCE and QCS. Preliminary versions of this work were presented at the conferences AQIS12 (Suzhou China, September 2012) and QANSAS11 (Agra India, December 2011).

\end{document}